\input{aipcheck}

\documentclass[
    ,final            
  ]
  {aipproc}

\layoutstyle{6x9}

\begin{document}

\title{$P_{11}$ Resonances with Dubna-Mainz-Taipei\\ Dynamical Model  for $\pi N$ Scattering and \\Pion Electromagnetic Production}


\author{Shin Nan Yang}{
  address={Department of Physics and Center for Theoretical
Sciences, \\National Taiwan University, Taipei 10617, Taiwan} }

\author{ S.S. Kamalov}{
  address={Bogoliubov Laboratory for Theoretical Physics, JINR,
Dubna, 141980 Moscow Region, Russia} }

\author{L. Tiator}{
  address={Institut f\"ur Kernphyik, Universit\"at Mainz, D-55099 Mainz,
Germany} }

\begin{abstract}
 We present the results on $P_{11}$ resonances obtained with
 Dubna-Mainz-Taipei (DMT) dynamical model for pion-nucleon scattering and
 pion electromagnetic production. The extracted values agree well,
 in general, with PDG values. One pole is found corresponding
 to the Roper resonance and two more resonances are definitely needed in
 DMT model. We further find indication for a narrow $P_{11}$
 resonance at around 1700 MeV with a width $\sim$ 50 MeV in both
 $\pi N$ and $\gamma \pi$ reactions.
\end{abstract}

\maketitle



  One of the most important tasks in the
  study of baryon structure is to extract their properties like,
  mass, width, and helicity amplitudes etc. from  $\pi N$
  scattering and pion electromagnetic (EM) production.
There are several approaches to  extract properties of nucleon
  resonances ($N^*$) from  $\pi N$ data, like speed plot,
  regularization method, dispersion relation,   and
  meson-exchange model etc.

  Interest on the properties of $P_{11}$ resonances  has recently
  intensified considerably \cite{Suzuki10,Kamano10} after the properties
  of $\Delta(1232)$ is well studied \cite{Pascalutsa07}. In this contribution,
  we present results on $P_{11}$ resonances obtained from analyzing
  data of $\pi N$  scattering and pion  EM
  production with Dubna-Mainz-Taipei (DMT) meson-exchange model which we
  have constructed in \cite{Chen03,Chen07}.

The  DMT $\pi N$ meson-exchange model was developed on the basis of
the Taipei-Argonne $\pi N$ meson-exchange model~\cite{Hung01} which
describes pion-nucleon scattering up to 400~MeV pion laboratory
energy. It starts from an effective chiral Lagrangian. The effective
Lagrangian is then used to construct a potential for use in the
scattering equation
\begin{equation}
t_{\pi N}=v_{\pi N}+v_{\pi N}g_0t_{\pi N}. \label{tpiN}
\end{equation} The Taipei-Argonne $\pi N$ model was
extended up to c.m. energies $W=2.0$~GeV \cite{Chen07} by inclusion
of higher resonances and the $\eta N$ channel, namely,
\begin{equation}
t_{ij}(W)= v_{ij}(W)+\sum_k  v_{ik}(W)\,g_k(W)\, t_{kj}(W)\,,
\label{tcoupled}
\end{equation}
with $i$ and $j$ denoting the $\pi$ and $\eta$ channels and $W$ is
the total c.m. energy. The potential $ v_{ij}$ is a sum of
background $v^B_{ij}$ and bare resonance $v^R_{ij}$ terms,
\begin{equation}
v_{ij}(W)=  v^B_{ij}(W)+  v^R_{ij}(W)\,, \label{vij}
\end{equation}
where
\begin{equation}v^R_{ij}(W)=\sum^n_{k=1} v^{R_k}_{ij}(W),\label{vR}
\end{equation} if there are $n$
resonances in any specific channel. The background term
$v^B_{\pi\pi}$ for the $\pi N$ elastic channel is taken as obtained
in \cite{Hung01}. The model describes well both $\pi N$ phase shifts
and inelasticity parameters in all the channels up to the $F$ waves
and $W= 2$ GeV, except for the $F_{17}$ partial wave.

The DMT $\pi N$ model is also a main ingredient of the DMT dynamical
model describing the photo- and electroproduction of pions
\cite{KY99} up to 2~GeV, which   can be expressed, in analogy to Eq.
(\ref{tpiN}), as
\begin{equation}
t_{\gamma\pi }=v_{\gamma\pi}+v_{\gamma\pi}g_0t_{\pi N},
\label{tgamapi}
\end{equation}
where $v_{\gamma\pi}$ is the transition potential which describes
the production of a pion by an incident photon and is a sum of
background  and resonance terms. The background term
$v^B_{\gamma\pi}$ is obtained from the same effective chiral
Lagrangian for $\pi N$ with the use of minimal substitution. The
resonance term $v^R_{\gamma\pi}$  also contains contribution from
the  $n$ resonances as appeared in Eq. (\ref{vR}) since all of them
can be excited by a photon. The details can be found in
\cite{Chen03,KY99}. This model gives excellent agreement with pion
production data from threshold to the first resonance
region~\cite{Kamalov00,Kamalov01}. For details, we refer readers to
\cite{Chen03,Chen07,Hung01}.

From Eqs. (\ref{tpiN}-\ref{tgamapi}), it is seen that the resonance
parameters extracted would be affected by the choice of background
terms $v^B_{ij}$ and $v^B_{\gamma\pi}$, which can best be tested in
the low-energy region where the resonance contributions are expected
to be negligible. It has been shown that the DMT model describes
well the $\pi^0$ photo- and electroproduction at threshold
\cite{Kamalov01}, namely, the $s$-wave multipoles $E_{0+}(\pi^0p)$
and $L_{0+}(\pi^0p)$, and the recent measurement of
electroproduction differential cross section at threshold with
$Q^2=0.05$ (GeV/c)$^2$ \cite{Weis08}. A new measurement on the
polarized linear photon asymmetry \cite{Hornidge10} which is found
to be very sensitive to the small $p$-wave multipoles, and the
extracted multipole $_pM_{1-}$ \cite{Tiator11} also agree well with
the DMT predictions. This nice agreement between data and DMT
predictions in the case of $p$-wave multipoles validates, to some
extent, the reasonablenss of the choice of $v^B_{ij}$ and
$v^B_{\gamma\pi}$ used in the extraction of the properties of
$P_{11}$ resonances from $\pi N$ scattering and pion  EM production
data with DMT model.

\begin{figure}
\includegraphics[width=0.34\linewidth,angle=90]{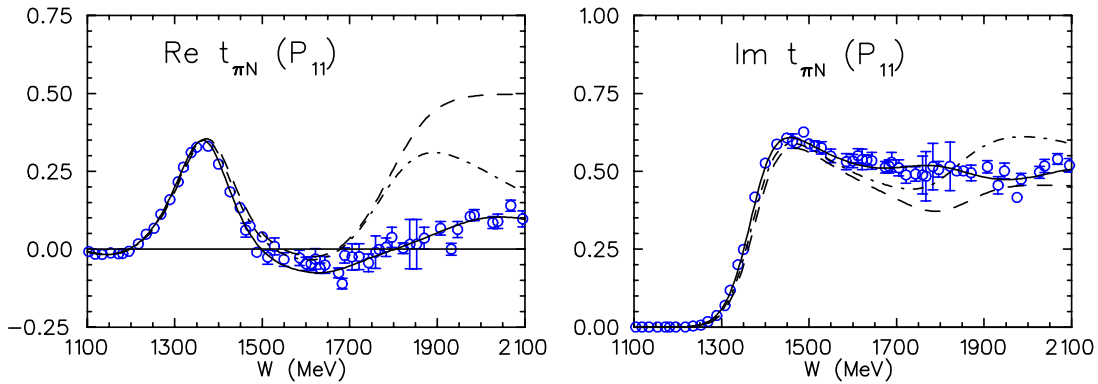}
\caption{Our predictions for $t_{\pi N}(P_{11})$. The The solid,
dashed, and dash-dotted curves   correspond to the predictions of
full DMT model, without the 2nd and 3rd resonances, and with the
removal only second resonance, respectively. The data are from
\cite{SAID}.}\label{tpiN-P11}
\end{figure}

The model parameters, including bare resonance masses, coupling
constants, and cut-off parameters for each resonance are then fitted
to $\pi N$ phase shifts and inelasticity parameters in all channels
up to $F$ waves and for energies less than 2 GeV \cite{SAID}. The
results for the $P_{11}$ channel is shown in Fig. \ref{tpiN-P11}.
The solid, dashed, and dash-dotted curves in Fig. \ref{tpiN-P11}
correspond to the predictions of full DMT model, without the 2nd and
3rd resonances, and with the removal only second resonance,
respectively. The indication for need of   three resonances in order
to reach a reasonable description of the data within DMT model seems
stronger than what was found by the EBAC group \cite{Kamano10}.

The bare masses, Breit-Wigner physical masses and widths, and the
pole positions extracted, all in units of MeV, are presented in
Table 1. The pole positions are determined by analytic continuation
and agree well with those obtained by speed plot and regularization
method \cite{Tiator10}. We find only one pole corresponding to the
Roper resonance, in contrast to the two poles found in
\cite{Suzuki10}. This is because we do not consider explicitly
$\pi\Delta$ channel and hence do not have a complex branch cut
associated with it. We did not try to look for poles beyond 2 GeV.

\begin{table}[hb]
\begin{tabular}{lccccc}
\hline
 $N^*$ & $M_R^{(0)}$ & $M_R$ & $\Gamma$ & Re$W_p$ &
 -Im$W_p$  \\
\hline
$P_{11}(1440)$  & 1612 & 1418 & 436 & 1371 & 95 \\
 PDG ****&  &$1445\pm 25$& $325\pm 25$ & $1365\pm 15$ & $95\pm 15$  \\
 \hline
$P_{11}(1710)$  & 1798 & 1803 & 508 & 1746 & 184 \\
 PDG ***   &  &$1710\pm 30$& $180\pm 100$ & $1720\pm 50$ & $115\pm 75$  \\
 \hline
$P_{11}(2100)$  & 2196 & 2247 & 1020&   &    \\
 PDG *  &  &$2125\pm 75$& $260\pm 100$ & $2120\pm240$ & $240\pm 80$  \\
 \hline
\end{tabular}
\caption{Bare $M^{(0)}_R$ and physical masses $M_R$, total widths
$\Gamma$, and the pole positions $W_p$, all in units of MeV
extracted for $P_{11}$ resonances with DMT model.}
\end{table}

With the inclusion of the three $P_{11}$ resonances determined from
$\pi N$ data as explained above, we are able to obtain a description
of the multipoles $_pM_{1-}(1/2)$ and $_nM_{1-}(1/2)$ in the  range
of photon energy $E_\gamma$ between  150-1650 MeV of comparable
quality obtained in the case $\pi N$ data as shown in Fig.
\ref{tpiN-P11}, except in the neighborhood of $E_\gamma\sim 1050$
MeV where a peak and a bump are seen in the imaginary and real part
of $_pM_{1-}(1/2)$, respectively. If a narrow $P_{11}$ resonance of
mass 1700 MeV and width of 47 MeV is included within MAID2007
\cite{MAID2007}, then the peak and bump mentioned above can be very
nicely described, as shown in Fig. 2.
\begin{figure}
\includegraphics[width=0.39\linewidth,angle=90]{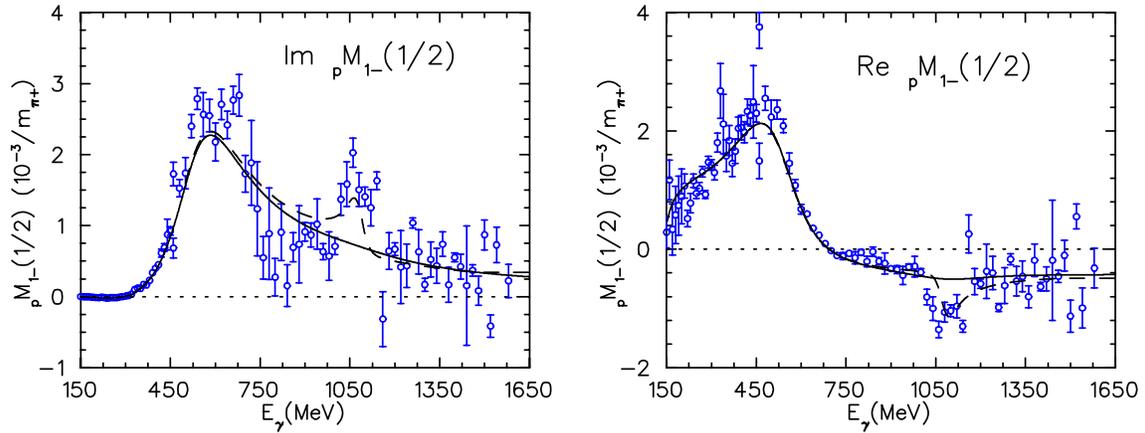}
\caption{Solid and dashed line denote results of MAID2007
\cite{MAID2007} and MAID2007 with an addition of a narrow $P_{11}$
resonance of mass 1700 MeV and width of 47 MeV, for $_pM_{1-}(1/2)$,
respectively.}
\end{figure}

In summary, we find, with the use of the DMT dynamical model for
analyzing the $\pi N$ scattering and pion EM production data, we
find three $P_{11}$ resonances at energies of 1418,  1803, and 2247
MeV, all with rather broad widths. We further find that, within MAID
2007, the inclusion of a narrow $P_{11}$ resonance at $W=1700$ MeV
with $\Gamma=47$ MeV can greatly improve the agreement with
$_pM_{1-}(1/2)$ data in the vicinity of $E_\gamma\sim 1050$ MeV.


\begin{theacknowledgments}
  The work of S.N.Y. is supported in part by the
NSC/ROC under grant No.~NSC99-2112-M002-011. We are also grateful
for the support by the Deutsche Forschungsgemeinschaft through
SFB~443, the joint project NSC/DFG 446 TAI113/10/0-3, and the joint
Russian-German Heisenberg-Landau program.
\end{theacknowledgments}


\bibliographystyle{aipproc}   


\end{document}